\documentclass[fleqn,10pt]{wlscirep}
\usepackage[utf8]{inputenc}
\usepackage[T1]{fontenc}

\usepackage{siunitx} 	%added by author
\usepackage{cleveref}	%added by author
\usepackage{caption}    %added by author
\usepackage{color,soul}  %added by author

\title{A pulsed optically pumped Rb clock with a frequency stability below $\mathbf{10^{-15}}$ }

\author[1]{Michele Gozzelino}
\author[1,*]{Salvatore Micalizio}
\author[1]{Claudio E. Calosso}
\author[2]{Jacopo Belfi}
\author[3]{Adalberto Sapia}
\author[4]{Marina Gioia}
\author[1]{Filippo Levi}
\affil[1]{Istituto Nazionale di Ricerca Metrologica, INRIM, Quantum Metrology and Nanotechnologies Division, Strada delle Cacce 91,  Torino, 10135, Italy}
\affil[2]{Space Engineering and Atomic Clocks,
Leonardo Electronics Società per Azioni,
Viale Europa snc, Nerviano (MI) 20014, Italy}
\affil[3]{
Space Engineering and Space Laser Systems,
Leonardo Electronics Società per Azioni,
Via dell’Industria 4, Pomezia (Roma),  00071, Italy}
\affil[4]{Optronics and Space Equipment LOB,
Leonardo Electronics Società per Azioni,
Viale Europa snc, Nerviano (MI), 20014, Italy}

\affil[*]{s.micalizio@inrim.it}

\begin{abstract}
We present the frequency stability performances of a vapor cell Rb clock based on the pulsed optically pumping (POP) technique. The clock has been developed in the frame of a collaboration between INRIM and Leonardo SpA, aiming to realize a space-qualified POP frequency standard. 
The results here reported were obtained with an engineered physics package, specifically designed for space applications, joint to laboratory-grade optics and electronics. %The clock physics package has been already fully implemented and characterized, whereas the electronics and the optical systems are still in the design phase. 
The measured frequency stability expressed in terms of Allan deviation is $1.2\times10^{-13}$ at 1s and achieves the value of $6\times10^{-16}$ for integration times of 40000 s (drift removed). This is, to our knowledge, a record result for a vapor-cell frequency standard. In the paper, we show that in order to get this result, a careful stabilization of microwave and laser pulses is required.

\end{abstract}
\begin{document}

\flushbottom
\maketitle
% * <john.hammersley@gmail.com> 2015-02-09T12:07:31.197Z:
%
%  Click the title above to edit the author information and abstract
%
\thispagestyle{empty}

%\noindent Please note: Abbreviations should be introduced at the first mention in the main text – no abbreviations lists. Suggested structure of main text (not enforced) is provided below.

\section*{Introduction}

Because of their reliability, compactness and good performances, vapor-cell clocks are nowadays employed in a large variety of scientific and technological applications that require precise timekeeping, jointly to reduced size, weight and power consumption (SWaP). It is sufficient to mention that global navigation satellite systems (GNSS), telecommunications, and timestamping of financial transactions all rely on precise time and frequency signals provided by atomic frequency standards which very often are Rb cell clocks \cite{camparo2007}.

\
Commonly used Rb clocks are lamp-pumped devices: a lamp is used as optical source for atomic state preparation through the process of optical pumping \cite{vanieraudoin}. However, since their introduction in atomic physics in the 80s, diode lasers have been successfully exploited in cell standards with the goal to improve the optical pumping process. In addition, due to the large number of available wavelengths, diode lasers allow the use of other atoms, like Cs, and are suited for implementing new excitation schemes, like coherent population trapping (CPT) (see for example \cite{yun2016, hafiz2015, godone2004,yun2017,Knappe2001}). 

At present, research on laser-pumped vapor cell clocks is an important and active field that roughly embraces two  trends: on one side the extreme miniaturization, aiming at realizing chip-scale clocks. On the other hand, the development of high-performing prototypes, with the goal of competing with H-maser clocks in terms of frequency stability, but achieving lower SwaP. 

In the first case, vapor-cell clocks as small as 1 cm$^3$ using a mm-scale cell have been demonstrated \cite{kitching2018}. If, on one hand, this miniaturization process shows many advantages (e.g. power consumption of a few tens of mW, reduced mass and production costs), on the other hand the short-term stability is necessarily limited to units of $10^{-10}$ at 1 s by the size of the microfabricated cell and then by the number of interacting alkali-metal atoms. Miniaturized atomic clocks have been shown to work successfully as time base for future GNSS receivers \cite{ramlall2011} and for seismic measurements related to earthquake detection, acoustic sensing, and oil exploration on the ocean floor \cite{gardner2012}. Also, miniaturized atomic clocks are developed in view of future applications in mobile and low-power instrumentation or hand-held devices \cite{tamagnin2021}.

The second research line concerns the development of laser-based vapor cell clocks with the highest stability performances. In this regard, several techniques have been devised and studied, adopting in most cases a cm-scale cell arrangement. These techniques include double-resonance continuous wave approach \cite{Almat2018}, pulsed optical pumping (POP) \cite{Almat2020, Micalizio2012, dong2017}, and CPT, either in continuous  \cite{yun2017, Yun2021} or in  pulsed regime \cite{hafiz2017, micalizio2019}. 
Among them, the POP scheme guarantees highly improved performances both with respect to current traditional Rb clocks and to competing new research ideas. After the seminal works based on the POP Rb maser \cite{godone2006}, it has been soon recognized that the optical detection of the ground state population makes it possible to achieve the best frequency stability results. Specifically, several research groups measured Allan deviations in the range from  $1\times 10^{-13}$ to $3 \times 10^{-13}$ for 1s of integration time. Also, in some cases, the medium-long term performances reached the low $10^{-15}$ region for $10^4$ s of averaging time \cite{Micalizio2012,Almat2020,Shen2020,Hao2020,Yun2021}.

In this paper, we report on the implementation and characterization of a Rb POP clock that exhibits an even better frequency stability result. In particular, an Allan deviation of $1.2\times10^{-13} \tau^{-1/2}$, being $\tau$ the integration time, has been measured. The white frequency noise region extends up to $4\times 10^4$ s (drift removed), allowing to reach the value $6\times10^{-16}$, a record result for a vapor cell standard. The clock prototype includes an engineered physics package and a laboratory prototype for the electronic and the optical systems. The Rb POP physics package (PP) has been designed for space applications in the frame of the ESA contract 'GSTP6.2 Rb POP' and has been characterized at INRIM's facilities from a physical and metrological point of view, including thermo-vacuum and magnetic-sensitivity tests \cite{Micalizio2021}. %Moreover, in view of its space employment, the physics package has been optimized in terms of size, volume and power consumption. 
Compared to the previous INRIM laboratory prototype \cite{Micalizio2012}, the PP has been engineered improving thermal behavior, shielding, and mechanical robustness to withstand space requirements, without losing in stability performances. Special effort in the design has been devoted to the reliability and repeatability of the assembly and tuning processes in view of a future series production. The PP has a mass of 3 kg and a power consumption less than 10 W.

The work is organized as follows. In section “The POP scheme”, for completeness, we lay out the pulsed approach and its advantages in the realization of a cell clock. In "Experimental Implementation" we describe the clock prototype and its operation. Particular emphasis will be given to the generation and control of the laser pulses.  “Results” section is devoted to the characterization measurements of the POP Rb clock, including the frequency stability results. Finally, in the conclusions, we summarize our results and outline some future perspectives.

\section*{The POP scheme}
The pulsed scheme allows the separation in time of the three phases (state preparation, clock interaction and detection) that usually characterize the operation of an atomic frequency standard. In traditional Rb clocks, preparation,  interrogation and detection take place at the same time: the light prepares the Rb atoms in one of two clock levels and simultaneously a microwave field resonant with the ground state hyperfine frequency is applied to the atomic sample in order to excite the clock transition. The same light used to pump the Rb atoms is also used for the detection stage, by measuring the excess of transparency induced by the microwave transition. This interrogation scheme, known as continuous-wave double-resonance, is quite efficient and is widely used in most commercial Rb clocks based on hot vapors.

However, the simultaneous presence of light and microwave generates a cross talk between optical and microwave transitions, resulting in a strong light-shift perturbing the clock frequency and then eventually affecting the stability of the reference signal \cite{Bandi2014}. Specifically, medium- and long-term performances are limited by the light shift, regardless of whether the clock uses the laser or the lamp for optical pumping. 

\begin{figure}[t]
\centering
\includegraphics[width=0.8\columnwidth]{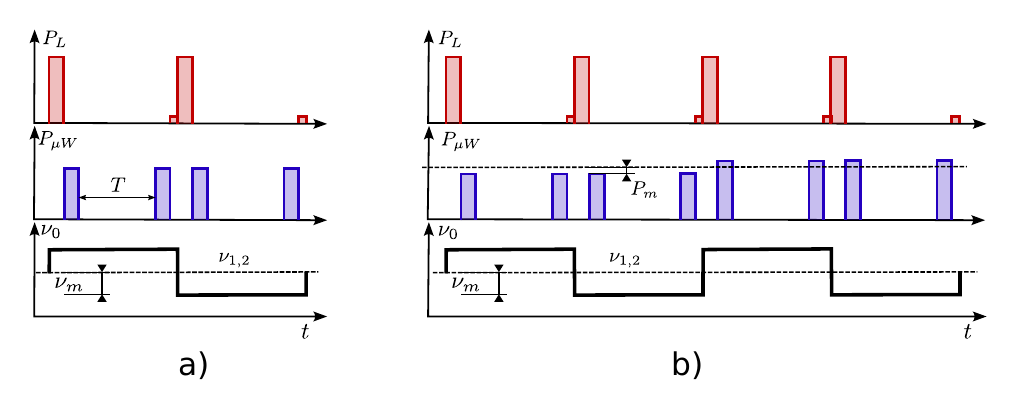}
\caption{a) Scheme of the basic POP clock cycle. $P_L$: laser power.  $P_{\mu W}$: microwave power. $\nu_0$: microwave frequency. Optical pumping and detection is performed with laser pulses (top plot). After the pumping phase, the laser is switched off and the clock interrogation is performed with a time-domain Ramsey sequence (middle plot). The sequence is repeated twice, modulating the microwave frequency $\nu_0$ with a modulation depth $\nu_m=\SI{80}{\hertz}$, i.e. half-width of the Ramsey fringe (bottom plot). The whole scheme lasts typically \SI{8.8}{\milli\second}. b) Extended sequence with an additional modulation of the microwave power (by $\pm\SI{8}{\percent}$) to perform an active stabilization of the microwave field amplitude on the atomic signal (see Methods).}
\label{fig:scheme}
\end{figure}

The pulsed optical pumping approach demonstrated very effective to reduce the light-shift. After being prepared in one clock level by an optical pumping pulse, the atoms make the transition in the dark stimulated by a couple of microwave pulses, according to the Ramsey scheme.  Differently from the double resonance continuous wave approach, the transfer of the laser instabilities to the atoms is minimized, light-shift is reduced by at least two orders of magnitude, with benefit for the medium-long term clock stability.

Without entering in all details, we remind here that a complete elimination of the light shift is not possible for two main reasons: first, it is  not possible to completely extinguish the laser light with standard opto-electronic devices and second, the laser absorption in the cell causes a non-homogeneity in the state preparation that results in a cell-position dependent shift \cite{English1978}.

Despite this so called pseudo-light shift effect, the POP technique shows some convenient features typical of a cold-atom fountain-like experiment. The atoms behavior can be well approximated by a two-level system, and the clock operation phases are well separated in time, being the Ramsey time no longer limited by gravity, like in a fountain, but by the relaxation phenomena taking place inside the cell. These mainly include the collisions with other Rb atoms (spin exchange), with the cell walls and with buffer gas atoms/molecules usually added to the cell to confine the atoms and to increase the coherence time \cite{vanieraudoin}.

The advantages of the POP scheme from a metrological point of view are not limited to the medium-long term. Indeed, during the optical pumping pulse, the laser intensity can be optimized to achieve the highest population inversion, improving the signal-to-noise ratio and consequently the short-term stability. Moreover, the central Ramsey fringe that represents the clock reference signal is nearly insensitive to any optical/microwave power broadening effect, being its linewidth (full-width at half maximum, $\Delta\nu_{1/2}$)  well approximated by the expression $\Delta\nu_{1/2}=\frac{1}{2T}$, where $T$ is the Ramsey time.

%Due to its effectiveness, the pulsed approach has been also successfully applied to CPT based vapor cell clocks, even if in this case the atoms cannot be assimilated to two-level systems.

A typical POP clock sequence is depicted in \Cref{fig:scheme}(a). The sequence starts with a strong resonant laser pulse creating a population imbalance in the two clock levels and almost extinguishing the residual atomic coherence from the previous cycle.  Then, the Ramsey spectroscopy is performed in the microwave domain. Finally, the ground state population is probed by a weak laser pulse (at the same frequency as the pump pulse). The basic sequence is repeated twice as the microwave frequency is modulated around the central Ramsey fringe ($\nu_m \simeq \Delta\nu_{1/2} / 2$).
In \Cref{fig:scheme}(b) an advanced interrogation scheme is employed to stabilize not only the local oscillator frequency, but also the microwave field amplitude (See Methods for more details).

\begin{figure}[hb]
\centering
\includegraphics[width=0.5\columnwidth]{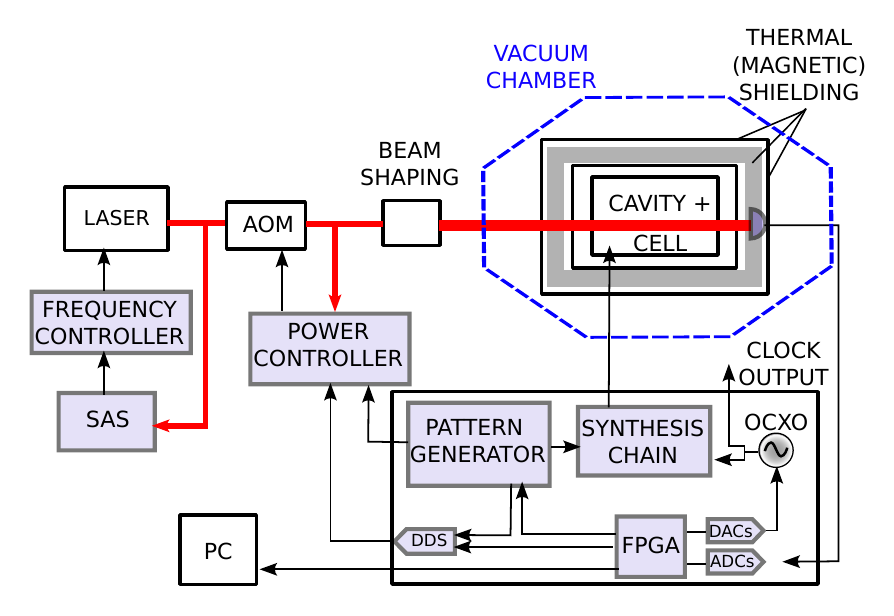}
\caption{Functional block scheme of the POP clock described in the text. For the frequency characterization, the engineered physics package was stored inside a vacuum chamber to provide a representative environment for on-ground tests.. SAS: Saturation Absorption Setup. FPGA: Field-Programmable Gate Array. DAC: Digital-to-Analog Converter. ADC: Analog-to-Digital Converter. AOM: Acousto-optic Modulator. OCXO: Oven-controlled Crystal Oscillator. }
\label{fig:blocks}
\end{figure}

\section*{Experimental implementation}

The experimental setup is shown in \Cref{fig:blocks}. Several relevant functions are performed by a dedicated digital electronic system that implements the synthesis chain starting from a 100 MHz quartz oscillator, the signal acquisition, and the signal processing for disciplining the quartz to the atomic reference. The clock interrogation cycle of \Cref{fig:scheme} is achieved by implementing a series of programmable steps that repeat themselves indefinitely \cite{Calosso2017}.

The Engineering Model (EM) tested in this work is a Physics Package (PP) fully representative in terms of  size, form and functionality to the flight model. The PP is equipped with not-screened components, identical to the flight counterparts, and it is assembled following space-qualified processes. The environmental tests on the PP EM included thermal-vacuum cycles, sinusoidal and random vibrations, and mechanical shock. They have been performed following the guidelines for the qualification of a full Galileo clock unit.

The PP is composed of a layered structure similar to that described in \cite{Micalizio2012}. The core of the PP is the cell that contains the Rb atoms and a temperature compensated mixture of buffer gases. The cell is housed in a microwave cavity which is fed with a signal resonant with the ground state hyperfine transition of $^{87}$Rb (6.834 GHz). A solenoid (for generating the quantization magnetic field), thermal and magnetic shields complete the PP. The main physical parameters, such as cell length, buffer gas content, operational temperature and cavity quality factor are very similar to those reported in \cite{Micalizio2012}.

Currently, two Rb POP PP EM units manufactured by Leonardo S.p.A have been thoroughly analysed and approved following a formal ESA review process and successfully passed the environmental (thermal and mechanical loads) and the performance test phases. In addition, a long-term test on one PP EM is ongoing since more than 3 years in INRIM facilities, confirming excellent stability of its critical parameters.

The optics package is composed of a distributed feedback (DFB) laser whose frequency is stabilized to the D$_2$ line (780 nm) via common saturated absorption spectroscopy (SAS) technique on the Rb $|\mathrm{F=1}\rangle \rightarrow |\mathrm{F'=1,2}\rangle$ crossover transition with \SI{-160}{\mega\hertz} offset, to compensate for the buffer-gas shift. A fiber-coupled acousto-optic modulator (AOM) acts both as a switch and as actuator for stabilizing the laser-pulses amplitude, according to the POP timing sequence. The pumping phase typically lasts \SI{0.4}{\milli\second} for a pump intensity of \SI[per-mode=symbol]{14}{\milli\watt\per\centi\meter\squared}. The detection pulse intensity is instead around \SI[per-mode=symbol]{0.9}{\milli\watt\per\centi\meter\squared}.  
Before entering the PP, the laser beam is shaped by a few free-space optics (see \Cref{fig:BB+RIN}a): after collimation, the linear polarization is adjusted with a half-wavelength plate and cleaned with a Glan-Thompson polarizer. Then, the beam is expanded to a $1/e^2$ beam diameter of \SI{6.6}{\milli\meter}. A non-polarizing beam-splitter is inserted just before the beam-expander to sample the laser intensity on a photodiode. The amplitude of the laser pulses is actively stabilized on this photodiode signal with 40 kHz bandwidth to improve the relative intensity noise (RIN); at the same time, the stabilization technique provides a control of the laser pulses area. More details can be found in Methods. 

\begin{figure}[!h]
\centering
\includegraphics[width=0.55\columnwidth]{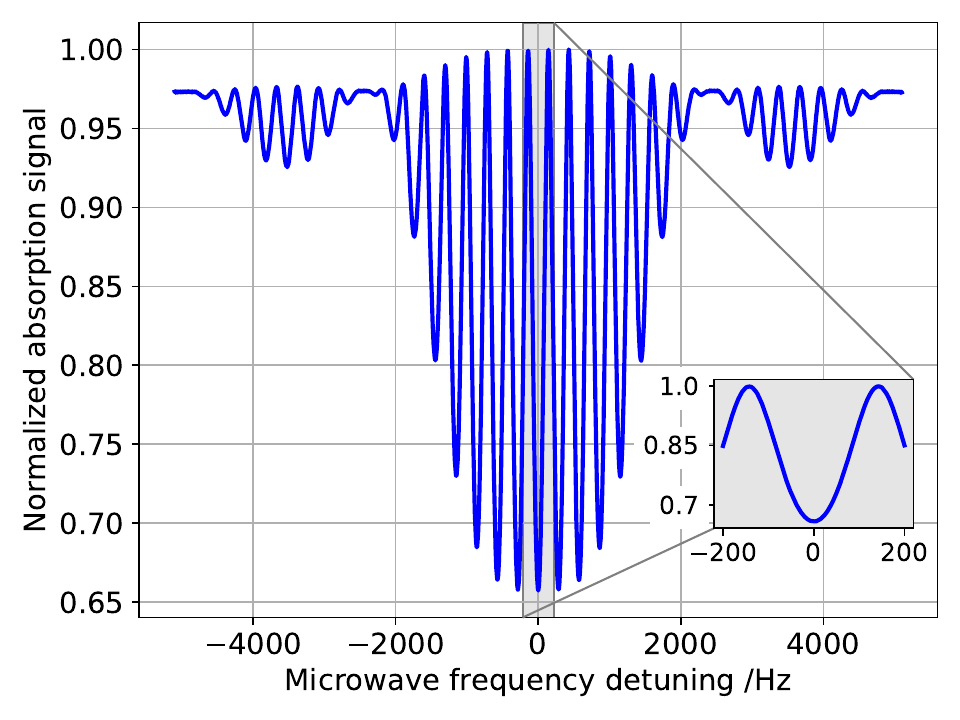}
\caption{Typical experimental Ramsey fringes measured with the engineered physics package. No averaging is made when scanning the local oscillator frequency (i.e. each data point is the result of one clock cycle).}
\label{fig:ramsey}
\end{figure}

\Cref{fig:blocks} shows a functional block-scheme of the whole clock, whereas \Cref{fig:BB+RIN}a presents a scheme of the optical breadboard, including the laser, the AOM, the beam-delivering system and the pulses-generation setup. The engineered
PP was stored inside a vacuum chamber, to provide a representative environment for testing. The typical clock signal achieved with the system is shown in \Cref{fig:ramsey}. Notably, the clock signal has similar or even better characteristics to the one obtained in the best-performing laboratory prototypes \cite{Almat2020, Micalizio2012}. 

\section*{Stability Results}

The stability results presented in this section refer to the latest extensive measurement campaign performed on one engineering model of physics package. 
%The physics package is kept under vacuum and has been characterized using optics and electronics packages developed at INRIM.
After an early evolution of the clock frequency due to the cell 'burn-in', the clock frequency behavior was rather predictable and shows a constant linear drift of $\simeq \SI[per-mode=symbol]{4e-14}{\per\day}$, even after a few programmed periods of "cold" (\SI{23}{\degreeCelsius}) and "hot" (\SI{65}{\degreeCelsius}) storage,  where the temperature controllers were not active and the package was left to thermalize with the environment. %The most reproducible operation was achieved with an active stabilization of both the laser and the microwave pulses.
\begin{figure}[!h]
\centering
\includegraphics[width=0.55\columnwidth]{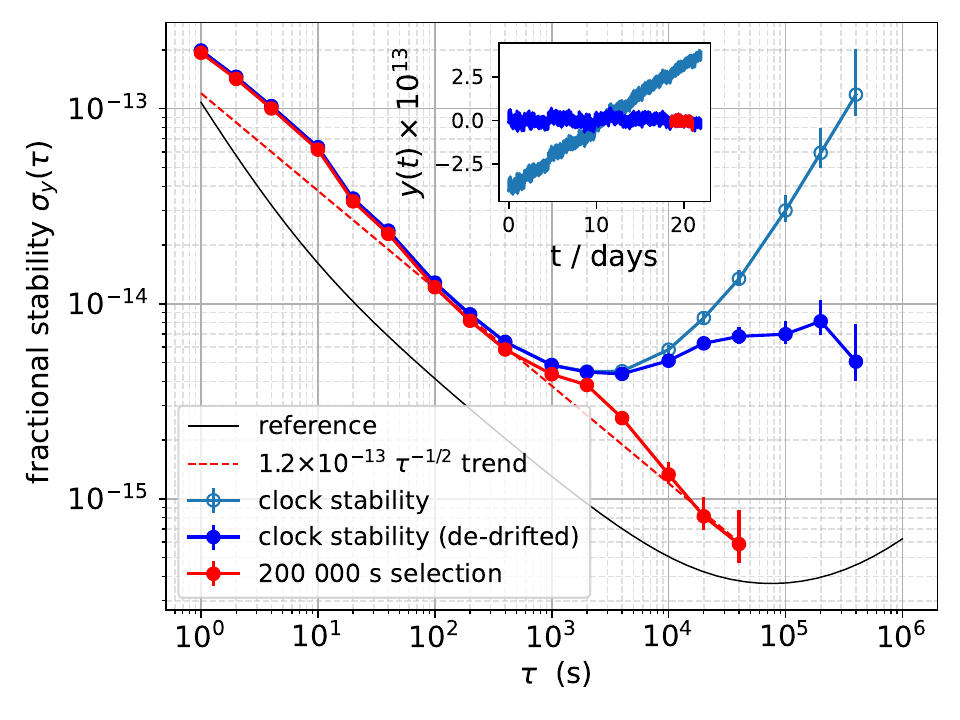}
\caption{Clock stability as measured versus an active hydrogen maser (measurement bandwidth 5 Hz). Light-blue open circles: overlapping Allan deviation (OADEV) of the whole measurement. Blue: same data with linear drift removed (\SI[retain-explicit-plus, per-mode=symbol]{+3.9e-14}{\per\day}). Red: selected sub-set (\SI{200000}{\second} long). Black: model of the active hydrogen maser used as reference. The inset shows the time series of the same data-sets with corresponding colors. In this case, data are averaged with a window of \SI{100}{\second} for better visualization.}
\label{fig:stability1}
\end{figure}

\Cref{fig:stability1} shows the result of a 22-days uninterrupted measurement of the Rb clock using an active hydrogen maser as a reference. The typical stability is \num{1.2e-13} $\tau^{-1/2}$ up to \SI{1000}{\s}, after that we observe a flicker noise in the mid $1\times10^{-15}$ region up to few-days averaging times. The small amount of excess noise in the short-term is due to spurs, partly arising from the clock signal, and partly from the distribution line of the hydrogen maser signal, and averages down faster than white noise as a function of the averaging time $\tau$. By selecting a \SI{200000}{\second} sub-set of the data, where the environmental parameters were more stable, we see the potential of the system to reach a stability region below \num[]{1e-15} after only \SI{1e4}{\second} of averaging time.

\begin{figure}[htpb]
     \centering
     \includegraphics[width=\textwidth]{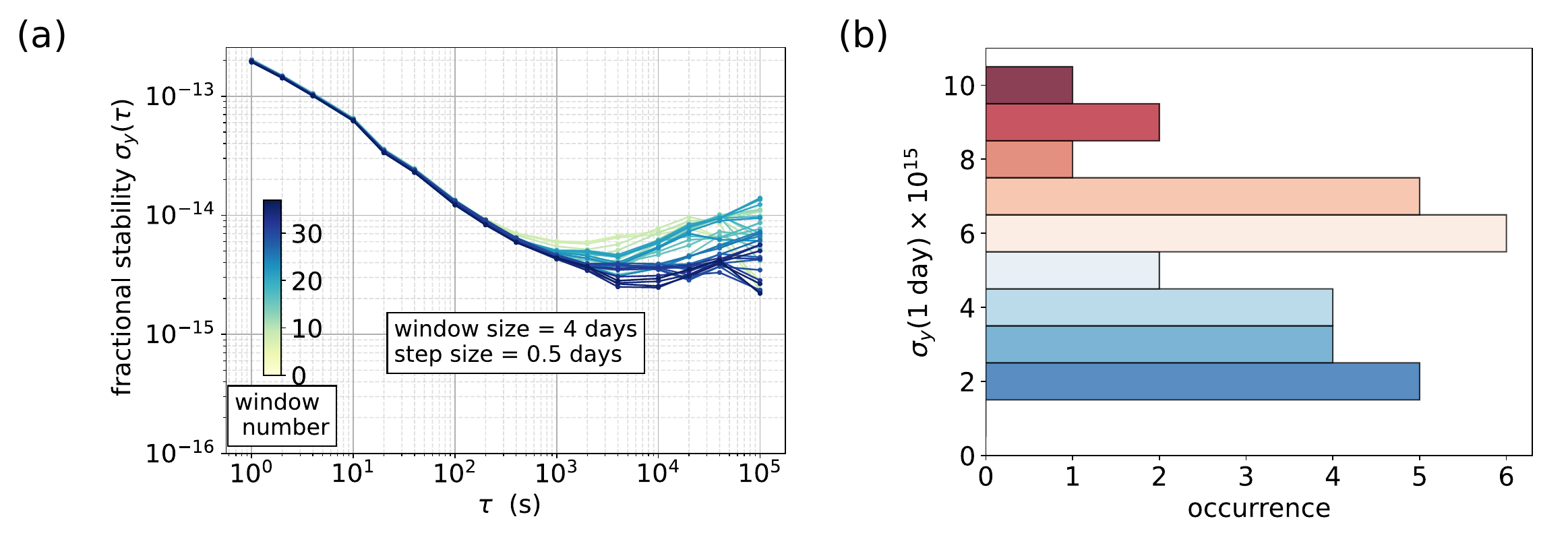}
     \caption{Analysis of the stability at one day averaging time for the same frequency data of \Cref{fig:stability1}. a) Dynamic Allan deviation (DAVAR) \cite{Galleani2005}: Allan deviation calculated over sliding time windows. Each window is 4-days long and translated from the preceding one by \SI{0.5}day. The color indicates the position of the window (starting with light-yellow at the beginning of the measurement and dark-blue at the end). b) Distribution of the stability at one-day averaging time.}
     \label{fig:DADEVs}
\end{figure}

To underline the potential of the POP clock architecture for satellite-based navigation systems \cite{Jaduszliwer2021}, we calculated the Dynamic Allan variance for the same data with a window of 4 days, in order to estimate the stability at one day, which is the typical interval between adjacent clock synchronizations in modern GNSS systems \cite{Jaduszliwer2021}. A global drift of \SI[retain-explicit-plus, per-mode=symbol]{+3.9e-14}{\per\day} is removed from the whole measurement before slicing. In \Cref{fig:DADEVs}a, the Dynamic Allan variance is plotted as a series of overlapped curves. The color indicates the position of the window (starting with light-yellow at the beginning of the measurement and dark-blue at the end). In \Cref{fig:DADEVs}b the distribution of the ADEV values calculated at 24 hours is shown. For the duration of the measurement, the one-day frequency stability is around \num{6e-15} with highest probability and seldom above \num{7e-15}. This level of frequency stability corresponds to a time-keeping better than \SI{0.5}{\nano\second} after one-day without synchronization, and, notably, without the need of updating the clock drift model, since the linear drift is stable over time. This is well within the desired level of performance for a GNSS on-board clock, bringing the User-Range-Error contribution from the clock to a negligible level (below \SI{20}{\cm} for one day of holdover time) \cite{Jaduszliwer2021}.

\section*{Discussion}
We have presented the frequency stability results obtained with a prototype of Rb vapor cell clock working in pulsed regime. The clock is composed of a physics package engineered for space operation, joint to laboratory electronics and optics. The measured Allan deviation is $1.2\times10^{-13} \tau^{-1/2}$, reaching the record result of $6\times 10^{-16}$ for an integration time of 40000 s (drift removed).
A complete engineered Rb POP clock targets a white frequency noise regime for an averaging time of the order of one day, as required by the ultimate stability goal for GNSS \cite{Jaduszliwer2021}.

Although the transfer of laser and microwave instabilities to the atoms is considerably reduced compared to other techniques, the control of the fields fluctuations must be handled also in the pulsed approach. The stability performances here reported have been obtained adopting specific techniques to reduce the clock frequency sensitivity  to laser and microwave power instabilities (see Methods for details). 
%We point out that the implementation of these techniques does not affect the clock SWaP; as explained in the next section, the microwave amplitude stabilization scheme results in a

The ongoing activities are focused into the development of fully integrated and qualified optical and electronics packages to be assembled with the already existing PP. The  results presented here represent a significant advance in vapor cell clock technology. In terms of SWaP, the final prototype is expected to have a volume of 17 liter, a weight of 10 kg and a power consumption of 60 W. These features, joint to the above reported stability performances, would make the Rb POP clock very attractive  to its possible deployment in GNSS or in any other space applications.

\section*{Methods}

\subsection*{Microwave amplitude stabilization}
The microwave power delivered to the physics package during the Rabi pulses is of the order of $-20$ dBm. The sensitivity of the fractional clock frequency $y$ to this parameter is $\Delta y/(\Delta P_{\mu W}/P_{\mu W}) = \SI[per-mode=symbol]{-7e-14}{\per\percent}$. Therefore, to achieve \num[]{1e-15} fractional stability, the microwave power delivered by the synthesis chain must be better than \SI{0.014}{\percent}. Achieving this level for long averaging times is not a trivial task both with the system free-running and with traditional active stabilization systems. To tackle this issue, we implemented the stabilization algorithm described in \cite{Gozzelino2018}. 

The amplitude of the Rabi pulses is modulated synchronously with the clock cycle around the setpoint that maximizes the atomic absorption (i.e. around the first Rabi oscillation maxima). The double modulation (microwave frequency and amplitude) is performed by doubling the usual clock pattern. An independent error signal is retrieved by processing the clock signal and the correction is applied to the AM modulation input of the synthesis chain. The modulation depth is rather small, of the order of \SI{8}{\percent}. Such a small modulation is possible thanks to the low-noise properties of the atomic signal and to the small bandwidth required for the loop (power fluctuations are temperature-related and typically start to kick-in at hundreds of seconds). Due to the small modulation depth and to the symmetry of the Ramsey fringes, the main clock loop is not perturbed and the short term stability is not affected. 

This scheme has been validated to reach $1\times10^{-5}$ fractional stability in the delivered microwave power at one day averaging time and is intrinsically robust, since the information is retrieved directly from the same atoms contributing to the frequency reference signal. Indeed, since the microwave field is sensed inside the microwave cavity, the control can also compensate for variations of the microwave field power due to time-varying coupling efficiency or cavity aging. Notably, this scheme requires very little additional hardware and does not introduce dead time in the clock operation. 

\subsection*{Laser pulses stabilization}
Although orders of magnitude less impacting than with CPT or CW vapor-cell clocks, laser power is still a significant parameter of influence affecting the long-term of the POP clock\cite{Micalizio2010}. In the previous high-performance prototype\cite{Micalizio2012}, the power stability of our free-space laser system was not such a critical factor, since temperature sensitivity of the physics package was the main limitation to the mid-term stability. With the current system, that includes some fiber paths and aims at even better performances in the long-term, an active stabilization of the laser pulses power was necessary. 

\begin{figure}[t]
\centering
\includegraphics[width=0.9\columnwidth]{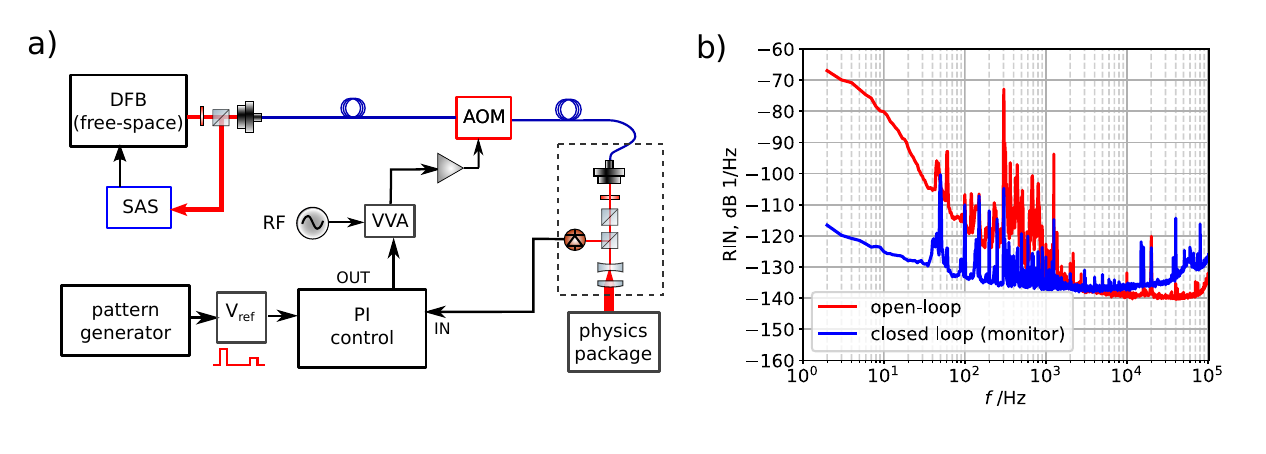}
\caption{a) Scheme of the optical breadboard, including the laser source, the beam-delivering system and the laser-pulses stabilization setup (see the text for a detailed explanation). DFB: Distribuited Feedback Laser, AOM: Acousto-Optic Modulator, VVA: Voltage Variable Attenuator, PI: Proportional-Integral, SAS: Saturation Absorption Setup. b) Relative intensity noise (RIN) of the laser as measured in CW with the control in open- and closed-loop. }
\label{fig:BB+RIN}
\end{figure}
For achieving the reported stability of \Cref{fig:stability1} we implemented a mixed analog/digital control to generate and stabilize the pulses. The laser power is probed just before the physics package by picking $\simeq \SI{30}{\percent}$ of the light with a non-polarizing beam splitter and acquiring it with a Si photodiode (see \Cref{fig:BB+RIN}a). The detected signal is conditioned and compared with a numerically controlled voltage reference. An analog proportional-integral (PI) controller acts on a variable voltage attenuator (VVA) that modulates the RF power sent to the single-pass AOM, controlling the amplitude of the laser and generating the pulses.

The voltage reference is composed of three programmable voltage levels (corresponding to "Pumping", "Detection" and "Dark" clock phases). The switching between the three states is performed with a fast analog switch (Analog Devices ADG412), driven by the same pattern generator of the clock electronics. The pattern also drives another switch that commutes between three different gains for the analog control. In this way, the gain during the three clock phases can be independently optimized. In this configuration, the controller is responsible for generating the laser pulses (by following the voltage reference) and reduces the laser RIN. The intensity noise achievable by the laser is determined by the voltage noise multiplied by the VVA+AOM trans-characteristic. The long-term stability of the pulse amplitude is instead stabilized (at best) at the level of the voltage reference. As seen in \Cref{fig:BB+RIN}b, the RIN reduction was between 10 dB and 20 dB in the \si{\kilo\hertz} range, achieving a white noise floor of $-135$ dB 1/Hz from \SI{100}{\hertz} up to \SI{50}{\kilo\hertz} (Fourier frequency). The long-term fractional power stability of the laser pulses power $\Delta P_L/P_L$ was of the order of \num{2e-4} (at 1 day) for both the detection and the pumping pulses.

\bibliography{biblio}

\noindent 

%For data citations of datasets uploaded to e.g. \emph{figshare}, please use the \verb|howpublished| option in the bib entry to specify the platform and the link, as in the \verb|Hao:gidmaps:2014| example in the sample bibliography file.

\section*{Acknowledgements}

This work has been partially funded under GSTP contract (Element 2) by the European Space Agency. The authors would like to thank Elio Bertacco for his invaluable help, especially in the design and optimization of the electronic controls.

\section*{Author contributions statement}

 M. Gozzelino, S. M., F.L. and C.E.C. performed the measurements and analysed the results. C.E.C. and M. Gozzelino designed and optimized the laser and microwave stabilization loops. S. M. and M. Gozzelino wrote the manuscript. M. Gioia, J. B., A. S. coordinated the Leonardo project. J. B., A. S. coordinated and contributed to the physics package design. All authors reviewed and contributed to the manuscript.

\section*{Additional information}

The authors declare no competing interests.

%To include, in this order: \textbf{Accession codes} (where applicable); \textbf{Competing interests} (mandatory statement). 

%The corresponding author is responsible for submitting a \href{http://www.nature.com/srep/policies/index.html#competing}{competing interests statement} on behalf of all authors of the paper. This statement must be included in the submitted article file.

\section*{Data availability}
The datasets used and/or analysed during the current study available from the corresponding author on reasonable request.

\end{document}